\documentclass[amssymb,amsmath,preprint,nofootinbib]{revtex4-1}
\usepackage[colorlinks]{hyperref}
\usepackage{graphicx}

\usepackage{gensymb}

\begin{document}
\title{Thermal leptogenesis in nonextensive cosmology}

\author{Mehran Dehpour}
\email{m.dehpour@mail.sbu.ac.ir}
\affiliation{Department of Physics, Shahid Beheshti University,
PO Box 19839-63113, Tehran, Iran}

\begin{abstract}
    Thermal leptogenesis is a mechanism that explains the observed asymmetry between matter and antimatter in the early universe. In this study, we review the impact of nonextensive Tsallis statistical mechanics on the early universe and study its effect on thermal leptogenesis. The study has found that the use of nonextensive statistical mechanics can affect the production of baryon asymmetry in thermal leptogenesis by modifying the equilibrium abundance of particles, decay, and washout parameters. Also, we show that nonextensive statistical mechanics potentially reduce the required right-handed neutrino mass scale.
\end{abstract}

\maketitle

\section{Introduction}
\label{sec:intro}
Observations suggest that there is an imbalance between the number of baryons (protons and neutrons) and antibaryons (antiprotons and antineutrons) in the universe. All the visible structures in the universe, such as stars, galaxies, and clusters, are made up of matter (baryons and electrons), and there is little to no antimatter (antibaryons and positrons) present. The baryon asymmetry of the universe is defined as
\begin{align}
    Y^{\rm obs}_{B} \equiv \left. \frac{n_B - \overline{n}_{B}}{s} \right|_0 = (8.73 \pm 0.35) \times 10^{-11},
\end{align}
where $n_B$, $\overline{n}_{B}$, and $s$ are the number densities of baryons, antibaryons, and entropy, respectively, and the subscript $0$ denotes the present time.
The amount of baryon asymmetry has been determined through Big Bang Nucleosynthesis (BBN), Cosmic Microwave Background (CMB), and Large Scale Structure (LSS) observations at a confidence level $95\%$~\cite{Simha:2008zj}.

If we consider universe was produced initially without baryon asymmetry or the initial baryon asymmetry was washed by inflation, the observed baryon asymmetry must have been dynamically generated. This process is called baryogenesis and it hinges on three essential components outlined by Sakharov in Ref.~\cite{Sakharov:1967dj}: violation of baryon number conservation, C and CP violation, and the presence of out-of-equilibrium dynamics. Notably, all these elements exist in the Standard Model (SM). However, it is important to note that within the SM framework, there is no mechanism capable of producing a sufficiently significant baryon asymmetry in~\cite{Gavela:1994ds,Gavela:1994dt}. Consequently, baryogenesis necessitates Beyond the Standard Model (BSM) physics, such as introducing new sources of out-of-equilibrium CP violation.

One of the possible BSM mechanisms for baryogenesis is thermal leptogenesis which was proposed by Fukugita and Yanagida in Ref.~\cite{Fukugita:1986hr}. New particles, right-handed neutrinos (RHNs), are introduced through the seesaw mechanism~\cite{Mohapatra:1980yp,Yanagida:1979as,Glashow:1979nm,Gell-Mann:1979vob,Minkowski:1977sc}. Their complex Yukawa couplings provide the new necessary source of CP violation. The drawback of thermal leptogenesis is the huge lower bound on RHN masses~\cite{Davidson:2002qv}. This bound could conflict with supersymmetric models because of the overproduction of gravitino~\cite{Kawasaki:2008qe,Rychkov:2007uq,Kawasaki:1994af,Khlopov:1984pf,Weinberg:1982zq}. The existence of this lower bound presents a significant phenomenological challenge, as the energy scale in question is currently beyond the reach of experimental exploration. However, modern developments in thermal leptogenesis aim to address these issues. A subset of these efforts involves the utilization of nonstandard cosmologies, such as those discussed in Refs.~\cite{Dehpour:2023wyy,sym12020300,Dutta:2018zkg,PhysRevD.90.064050}.

In this study, we focus on employing a nonstandard cosmology development approach by modifying statistical mechanics. Recent research has demonstrated that conventional statistical mechanics is not universally applicable, as the statistical distribution of some systems has power-law tails instead of the usual exponential. In 1988, Tsallis introduced a generalized framework known as nonextensive statistical mechanics~\cite{Tsallis:1999nq,Tsallis:1987eu}. The deviation from standard statistical mechanics, encoded in the Tsallis parameter $q$, is recovered when $q=1$.
Currently, it is not clear what causes power-law tails. However, in certain instances, the index $q$ can be determined analytically based on microscopic quantities, as well as other indices whose comprehensive list is detailed in Refs.~\cite{tsallis_introduction_2023}.
Especially, it is widely recognized that high energy effects, including strong interactions and microscopic fluctuations, can exhibit power-law tails behaviors in statistical distributions~\cite{Biro:2008hz,Alberico:2009gj,Wilk:2008ue,Kodama:2009hme,Kaniadakis_2009,Beck:2009uy}.
Consequently, it has been proposed that the early universe might also exhibit similar power-law tails behaviors. However, this paper explores the effects of nonextensive statistical mechanics in the early universe without delving into its origins.
Recently, the nonextensive cosmology has been employed to describe BBN~\cite{Hou:2017uap,Bertulani:2012sv} and dark matter production~\cite{Rueter:2019ubf}.\footnote{The exploration of nonextensive cosmology through considering the entropic origin of gravity and the modification of gravity by altering statistical mechanics is another area of study~\cite{Sheykhi:2018dpn}. This approach to nonextensive cosmology has also been applied in the early universe, for instance, to describe the BBN~\cite{Jizba:2023fkp}.}

We propose a new method for studying thermal leptogenesis that considers nonextensive Tsallis statistical mechanics in the description of particle distributions. 
Our considered thermal leptogenesis model has three RHNs while for simplicity only the lightest one can decay. This affects the equilibrium abundance of particles and the Hubble expansion rate, resulting in modified decay and washout parameters involved in thermal leptogenesis. Our results indicate that, depending on whether $q>1$ or $q<1$, thermal leptogenesis in a nonextensive universe can generate less or more asymmetry than the standard. Furthermore, we note that greater production of asymmetry implies a reduction in the RHN mass scale.

The remainder of this paper is organized as follows. In Sect. \ref{sec:tsallis}, we briefly introduce nonextensive Tsallis statistics mechanics and review its effects on cosmology. In Sect. \ref{sec:modified}, we provide a detailed review of thermal leptogenesis and impose the nonextensivity effect on it. In Sect. \ref{sec:results}, we introduce the free parameters of the model and extract our results by numerically solving the Boltzmann equations. Finally, in Sect. \ref{sec:conclu}, we present discussions and conclusions.

\section{Nonextensive cosmology}
\label{sec:tsallis}
Before we describe the effects of nonextensive Tsallis statistical mechanics on cosmology, we describe the necessary ingredients in the Tsallis framework. The $q$-exponential function of $x$ is defined as~\cite{tsallis_introduction_2023}
\begin{align}
    e_q^x \equiv \left[1+\left(1-q\right) x\right]^{1/(1-q)},
\end{align}
where in the limit $q \to 1$, $e_q^x \to e^x$.

By generalized definition of entropy, the generalized distribution function is parameterized by a real number $q\in[0,2]$ known as Tsallis parameter~\cite{tsallis_introduction_2023}
\begin{align}
    f^{q} = \frac{1}{e_q^{(E-\mu)/T}+\xi},
    \label{eq:dist}
\end{align}
where $T$, $\mu$, and $E$ denote the temperature, chemical potential, and energy, respectively. In addition, $\xi$ is equal to $0, -1\ \text{or}\ 1$ for a Maxwell-Boltzmann (MB), Bose-Einstein (BE), and Fermi-Dirac (FD) gas.
In this way, for $q>1$, distribution has a fat tails while for $q<1$, it has sharp cutoff tails.
We define $e_q^x \equiv 0$ in two cases: (i) $q<1$ and $x < 1/(q-1)$ and (ii) $q>1$ and $x \ge 1/(q-1)$, respectively, which are interpreted as cutoff distribution functions at high energies $E\ge \mu - T/(q-1)$ and low energies $E \le \mu - T/(q-1)$.

It is important to note that since properties of particles are different from each other and evolutionary over time, $q$ for each particle can be different and evolve. However, in this study, we choose the simplest one in which $q$ is equal for all particles and fixed.

In this way, although the Einstein equations are considered fundamental and the Friedmann equations are not modified, when the particle distributions are altered, the energy density of matter which is proportional to the integral of the distribution function, is also changed. Consequently, the Hubble expansion rate must be revised. Thus, as we work in the early universe, in the radiation-dominated phase, the modified Hubble expansion rate in nonextensive cosmology given by~\cite{Pessah:2001mz}
\begin{align}
    H^q = \frac{1.66}{M_{Pl}} (g_{\star}^q)^{1/2} T^2,
    \label{eq:Hubble}
\end{align}
where $M_{\rm Pl} = 1.22 \times 10^{19}$ is Planck mass and $g_{\star}^q$ is the effective degree of freedom for energy density, as we work before the electroweak phase transition for massless particles is given by~\cite{Rueter:2019ubf}
\begin{align}
    g_{\star}^q &= \left[ \frac{15}{\pi^4} \int_0^{\infty} d\gamma \gamma^3 \left(e_q^{\gamma}-1\right)^{-q} \right] \sum_b g_b \notag \\
    &+ \left[ \frac{15}{\pi^4}\int_0^{\infty} d\gamma \gamma^3 \left(e_q^{\gamma}+1\right)^{-q} \right] \sum_f g_f,
\end{align}
Accoriding to $g_{\star}^q$, for $q<1$ opposite of $q>1$, Hubble rate could decreased. Thus, this modification makes the time for the reactions to reach thermal equilibrium earlier and one can expect this could affect leptogenesis. If the decay of RHN reaches equilibrium earlier, CP violation can increase.

Moreover, like energy density modification because of altered particle distributions, entropy density is changed too. One can calculate the modified entropy density in nonextensive cosmology during the radiation-dominated phase as~\cite{Pessah:2001mz}
\begin{align}
    s^q = \frac{2 \pi^2}{45} g_{\star, s}^q T^3,
    \label{eq:entropy}
\end{align} 
where $g_{\star, s}^q$, the  entropy density degrees of freedom, which for massless particles is~\cite{Rueter:2019ubf}
\begin{align}
    g_{\star, s}^q &= \left[ \frac{45}{4 \pi^4} \int_1^{\infty} d\gamma \left(\frac{4}{3}\gamma^3+\frac{\sqrt{\gamma^2-1}}{3} \right) \left(e_q^{\gamma}-1\right)^{-q} \right] \sum_b g_b \notag\\
    &+\left[ \frac{45}{4 \pi^4} \int_1^{\infty} d\gamma \left(\frac{4}{3}\gamma^3+\frac{\sqrt{\gamma^2-1}}{3} \right) \left(e_q^{\gamma}+1\right)^{-q} \right] \sum_f g_f.
\end{align}
Note that $g_b$ and $g_f$ are the boson and fermion degrees of freedom at a specific temperature, respectively. As during the radiation-dominated period, all particles are relativistic, $\sum_{f} g_f = 90$ and $\sum_{b} g_b = 28$~\cite{Husdal:2016haj}.

\section{Modified thermal leptogenesis}
\label{sec:modified}
Thermal leptogenesis is based on the concept of introducing RHNs, which interact with standard model particles via the Yukawa interactions and gravity. Similar to other standard model particles, these heavy sterile particles can be created through thermal mechanisms in the early universe. They can violate CP in their out-of-equilibrium decay through the Yukawa channel if the Yukawa couplings are complex, shown as soon as later in Eq.~(\ref{eq:CP-parameter}). This new source of CP violations leads to asymmetry. This asymmetry is communicated from singlet neutrinos to ordinary leptons through their Yukawa couplings. The lepton asymmetry is then reprocessed into baryon asymmetry by the electroweak sphalerons. 

Here, we consider a simple model in which three RHNs exist to solve the small mass of three active neutrinos through the seesaw mechanism while for simplicity just the lightest of them, $N_1$, can decay through the Yukawa interaction.\footnote{We need at least two RHNs as we are sure that at least two active neutrinos are massive and it is necessary to violate CP in loop corrections of RHN decay via Yukawa interaction which will be revealed in Eq.~(\ref{eq:CP-parameter}). It may be discovered in the future that the third active neutrino is massive, so we considered three RHNs.} The reactions involving $N_1$ can be described as
\begin{align}
    N_1 \rightleftarrows \bar{\phi} l,
    \label{eq:decay}\\
    N_1 \rightleftarrows \phi \bar{l}.
    \label{eq:decay-anti}
\end{align}
To quantify this scenario, one can first calculate the tree-level decay rates by definition of $\overline{\Gamma} \equiv \Gamma(N_1 \to \phi \bar{l})$, and $\Gamma_1 \equiv \Gamma(N_1 \to \bar{\phi} l)$~\cite{Davidson:2008bu}
\begin{align}
    \Gamma_1= \overline{\Gamma}_1 = \frac{M_1}{16 \pi} (yy^{\dagger})_{11},
    \label{eq:decay-rate}
\end{align}
where $M_1$ is the mass of $N_1$ and $y$ is the Yukawa coupling matrix. Note that the rates of these decays were lower than the Hubble rate at high temperatures.

When the temperature is lower than the mass of the lightest RHN, $M_1$, the reactions described in Eqs.~(\ref{eq:decay}) and (\ref{eq:decay-anti}) proceed only in one direction, from left to right. If the rates of these decay reactions are not equal, CP violation occurs.
We can then proceed to introduce a CP violation parameter that is adjusted to the total decay rate, which can be expressed as~\cite{Davidson:2008bu}
\begin{align}
    \epsilon_1  \equiv \frac{\Gamma_1 - \overline{\Gamma}_1}{\Gamma_1 + \overline{\Gamma}_1}.
\end{align}

The CP violation parameter is nonzero only if the loop corrections are taken into account,
\begin{align}
    \epsilon_1 = \sum_{k\neq1} \frac{1}{8\pi} \frac{\Im \left(yy^{\dagger}\right)_{1k}^2}{\left(yy^{\dagger}\right)_{11}} \left[ f\left(\frac{M_k^2}{M_1^2}\right) + \frac{M_1 M_k}{M_1^2 - M_k^2}\right],
    \label{eq:CP-parameter}
\end{align}
which in that
\begin{align}
    f(x) = \sqrt{x} \left[1-\left(1+x\right)\ln\left(\frac{1+x}{x}\right)\right].
\end{align}
According to Eq.~(\ref{eq:CP-parameter}), one can see that, it is necessary to have at least two RHNs for nonzero CP violation parameter.
Note that CP violation can increase if other RHNs decay in the Yukawa interaction.

The dynamics of the lepton asymmetry and the number density of RHN during their evolution is a non-equilibrium process that can be mathematically described using the Boltzmann equations in the Friedmann-Lemaitre-Robertson-Walker (FLRW) universe. Note that the main form of the Boltzmann equations is unchanged during the modification of statistical mechanics, as in Ref.~\cite{Pessah:2001mz} mentioned. So one can obtain~\cite{Buchmuller:2004nz,Davidson:2008bu}
\begin{align}
    \frac{dY^q_{N_1}}{dz} &= - D^q_1 \left( Y^q_{N_1} - Y_{N_1}^{{\rm eq}, q} \right),
    \label{eq:YN1}\\
    \frac{dY^q_{B-L}}{dz} &= - \epsilon_1 D^q_1 \left( Y^q_{N_1} - Y_{N_1}^{{\rm eq}, q} \right) - W^q_1 Y^q_{B-L},
    \label{eq:YBL}
\end{align}
where $z\equiv M_1/T$ is dimensionless parameter, $Y^q_{N_1}\equiv n^q_{N_1}/s^q$ is normalized RHN number density, and $Y^q_{B-L}=(\overline{n}_l^q -n^q_{l})/s^q$ is lepton asymmetry.
In the above equations, $Y_{N_1}^{{\rm eq},q}$ is the equilibrium value of the number density of RHN, $D^q_1$ is the decay parameter and $W^q_1$ is the washout parameter. 
The decay parameter refers to the reaction of Eqs.~(\ref{eq:decay}) and (\ref{eq:decay-anti}) proceeding from left to right, leading to the decay of RHN and producing of lepton and lepton asymmetry if $\epsilon_1$ nonzero, while the washout parameter is the opposite.
These will be derived and discussed in more detail later.

\subsection{Equilibrium amount of particles}
\label{sec:YEq}
\begin{figure}[h]
    \includegraphics[width=140mm]{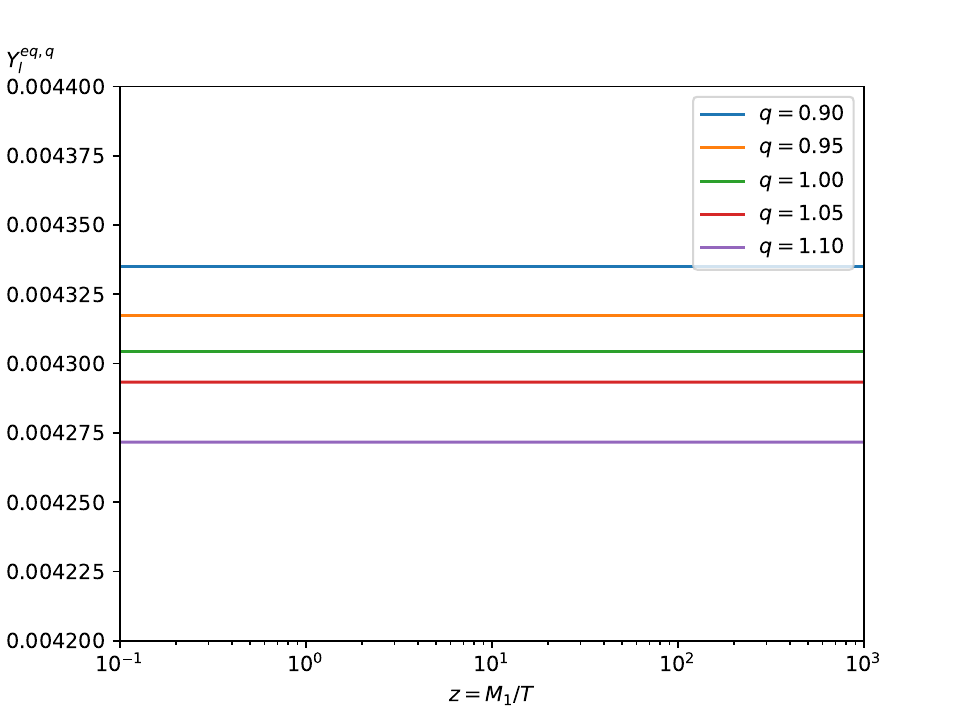}
    \caption{The equilibrium abundance of $l$ for some $q$ values with $M_1 = 10^{11}\ \rm GeV$ \label{fig:YEql}}
\end{figure}

\begin{figure}[h]
    \includegraphics[width=140mm]{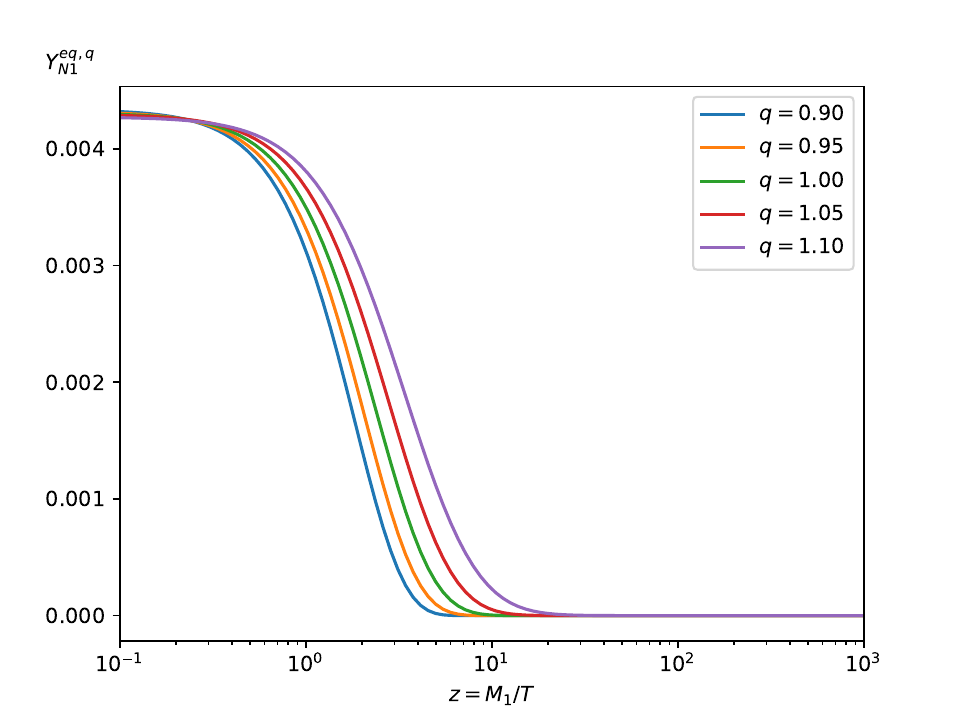}
    \caption{The equilibrium abundance of $N_1$ for some $q$ values with $M_1 = 10^{11}\ \rm GeV$ \label{fig:YEqN1}}
\end{figure}

One can find the equilibrium number density of $\chi$ particle given by~\cite{Kolb:1990vq}
\begin{align}
    n_{\chi}^{{\rm eq},q} = g_{\chi} \int \frac{d^3 p}{(2 \pi)^3} f_{\chi}^{{\rm eq},q},
    \label{eq:def-n}
\end{align}
where $g_{\chi}$ is the degree of freedom of $\chi$ particle at a specific temperature and the modified nonextensive distribution function in Eq.~(\ref{eq:dist}) with $\mu = 0$ was used for $f_{\chi}^{{\rm eq},q}$. Moreover, the equilibrium number density is converted to $Y_{\chi}^{{\rm eq},q} = n_{\chi}^{{\rm eq},q}/s^q$, where $s_q$ is given by Eq.~(\ref{eq:entropy}). So, one can obtain,
\begin{align}
    Y_{\chi}^{{\rm eq},q} &= \frac{45}{4 \pi^4} \frac{g_{\chi}}{g_{\star,s}^q} \frac{z^3}{M_1^3} \int_{0}^{\infty} dp\ p^2 \left[ e_q^{(\frac{E_{\chi} z}{M_1})}+\xi_{\chi}\right]^{-1}.
    \label{eq:YEq}
\end{align}
To obtain $Y_{N_1}^{{\rm eq},q}$ we must simply substitute $g_{N_1}=2$ and $\xi_{N_1}=0$ as $N_1$ follows the MB distribution function because it has a huge mass and the density is small enough that quantum corrections are negligible. Similarly, to obtain $Y_{l}^{{\rm eq},q}$ we must substitute $g_{l}=2$ and $\xi_{l}=1$ because $l$ follows the FD distribution function.
In Figs.\ \ref{fig:YEql} and \ref{fig:YEqN1} we ploted $Y_{N_1}^{{\rm eq},q}$ and $Y_{l}^{{\rm eq},q}$. 
As shown in Figs. $Y^{{\rm eq},q}_{l}$ and $Y^{{\rm eq},q}_{N_1}$, for small values of $z$, are larger than the standard values for $q<1$, despite the cases $q>1$. 
Furthermore, by evolving while passing $z$, $Y^{{\rm eq},q}_{l}$ becomes constant at high temperatures, whereas $Y^{{\rm eq},q}_{N_1}$ starts to decrease. As the value of $q$ increased, the rate of decrease for $Y^{{\rm eq},q}_{N_1}$ was found to accelerate. As a result, even though the value of $Y^{{\rm eq},q}_{N_1}$ for $q<1$ was initially greater than that for $q>1$, this trend reverses as the value of $z$ increases.

\subsection{Decay parameter}
\label{sec:D1}
The decay parameter in the Boltzmann equations is defined by the expression~\cite{Buchmuller:2004nz}
\begin{align}
    D^q_1 \equiv  \frac{2}{H^q z} \langle \Gamma_{1} \rangle,
    \label{eq:def-D1}
\end{align}
where $\langle \dots \rangle$ denotes the thermal average. 
We write the thermal average of the decay rate which is obtained in Eq.~(\ref{eq:decay-rate}) as
\begin{align}
    \langle \Gamma_{1} \rangle = \langle \overline{\Gamma}_1 \rangle =  \langle\frac{M_1}{E_{N_1}}\rangle \frac{M_1}{16 \pi} (yy^{\dagger})_{11}.
    \label{eq:averaged-decay-rate}
\end{align}
Now one can calculate the thermal average of $1/E_{N_1}$ with MB distribution of $N_1$. So, the Eq.~(\ref{eq:def-D1}), can be written as
\begin{align}
    D^q_1= \frac{2}{H^q z} \frac{\int_{0}^{\infty} \frac{dp\ p^2}{E} e_q^{-(\frac{E_{N_1} z}{M_1})}}{\int_{0}^{\infty} dp\ p^2 e_q^{-(\frac{E_{N_1} z}{M_1})}} \frac{M_1^2}{16 \pi} (yy^{\dagger})_{11}.
    \label{eq:D1}
\end{align}
where the Eq.~(\ref{eq:Hubble}) can be substituted for the Hubble rate. It is clear, in this way, decay parameter must be a function of $q$.

\begin{figure}[h]
    \includegraphics[width=140mm]{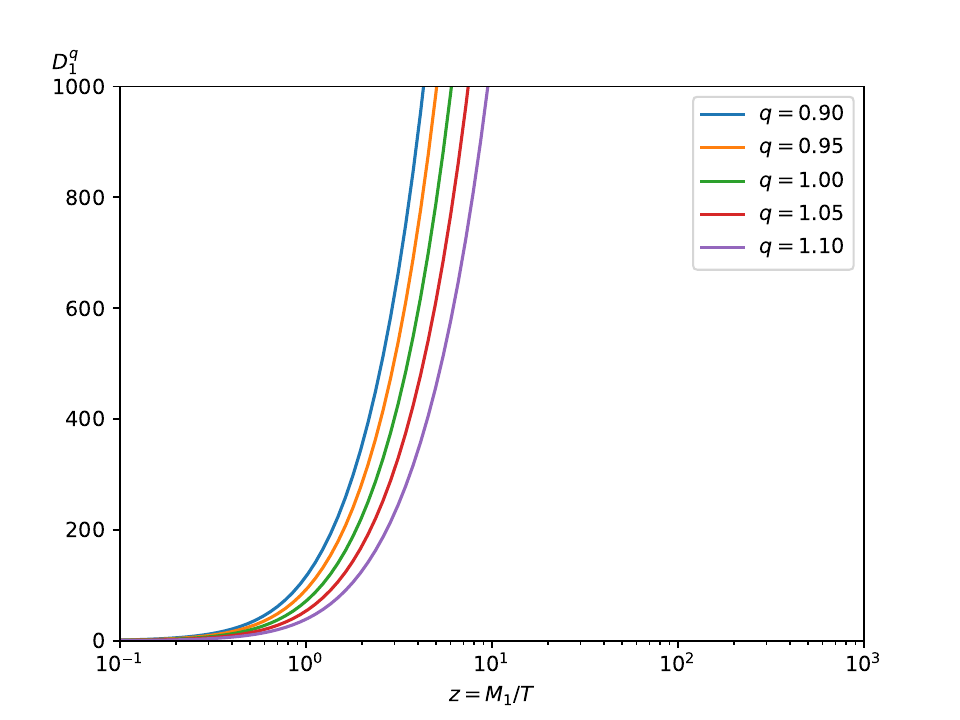}
    \caption{The decay parameters for some $q$ values with $M_1 = 10^{11}\ \rm GeV$ \label{fig:D1}}
\end{figure}

At first, the Hubble rate is larger than the RHN decay rate, the decay of RHN is negligible because it is out-of-equilibrium. However, by expanding the universe, the Hubble parameter decreases more quickly than the RHN decay rate and this reaction becomes increasingly effective and eventually attains thermal equilibrium.
According to Eq.~(\ref{eq:D1}) we plot the decay parameter for some $q$ values in Fig.\ \ref{fig:D1}. As one can see, when $q<1$, the decay parameter increases sooner than when $q>1$.
This implies that the generation of asymmetry for cases $q>1$ occurs later than in the standard case, unlike in $q<1$. However, because we work at very high temperatures, this does not affect the asymmetry near the electroweak phase transition.

\subsection{Washout parameter}
\label{sec:W1}
In Boltzmann equations, the washout parameter, $W_1$, is defined by the expression~\cite{Buchmuller:2004nz}
\begin{align}
    W^q_1 \equiv \frac{1}{2} \frac{Y_{N_1}^{{\rm eq},q}}{Y_{l}^{{\rm eq},q}} D^q_1,
    \label{eq:W1}
\end{align}
where $Y^{{\rm eq},q}_{N_1}$ and $Y^{{\rm eq},q}_l$ are described in Sect.\ \ref{sec:YEq} and $D_1^q$ is explained in Sect.\ \ref{sec:D1}.
As previously shown the decay parameter and equilibrium amount of $N_1$ are as functions of $z$, so the washout parameter also must be as a function of $z$.

\begin{figure}[h]
    \includegraphics[width=140mm]{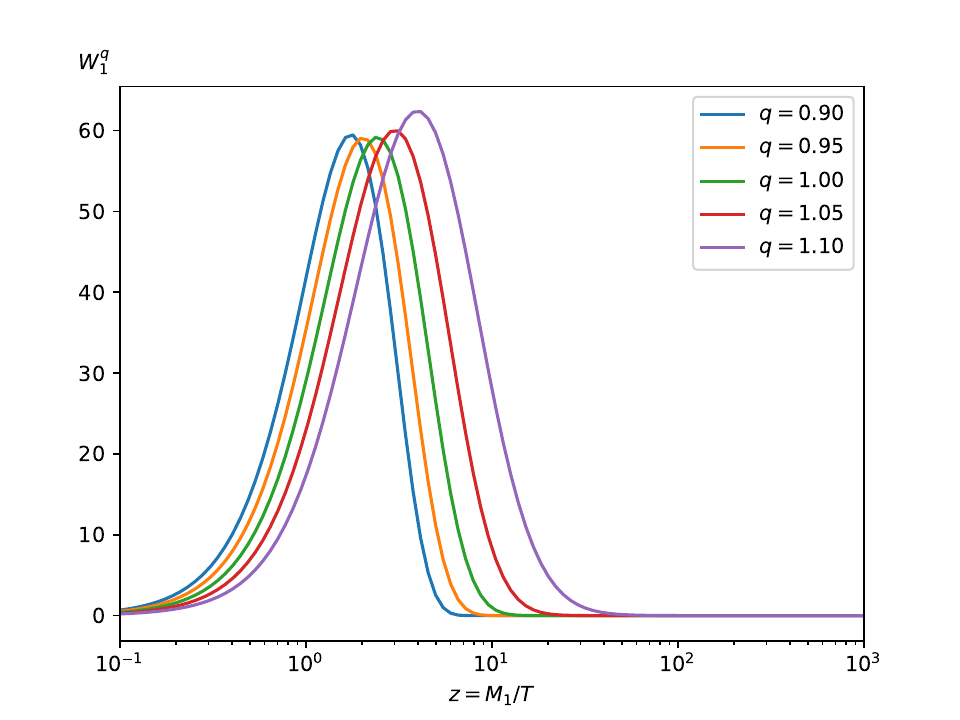}
    \caption{The washout parameters for some $q$ values with $M_1 = 10^{11}\ \rm GeV$ \label{fig:W1}}
\end{figure}

The reactions of Eqs.~(\ref{eq:decay}) and (\ref{eq:decay-anti}) cannot occur from right to left till the decay of RHN can produce a few lepton. Also, after a certain temperature, it cannot occur as there is not sufficient energy to produce a huge massive RHN.
We can plot the washout parameters for certain values of $q$. 
As depicted in Fig. \ref{fig:W1}, for values of $q>1$, the washout parameter attains its maximum at large $z$ values, in contrast to the situation for $q<1$. 
This behavior is expected due to the multiplication of the decay parameter curve by the equilibrium abundance of $N_1$ curve which are shown in Figs.~\ref{fig:D1} and \ref{fig:YEqN1}.
This has a significant impact on the produced asymmetry, as when the maximum washout parameter arises at small $z$ values, where RHN is not sufficiently produced, and consequently, CP violation is limited, the entire generated asymmetry will be erased. Thus, we can infer that for $q<1$ values, there will be more CP violations and a substantial amount of asymmetry.
Additionally, since the washout parameter has a division in the equilibrium abundance of $l$, according to Fig.~\ref{fig:YEql}, as it decreases for $q>1$, the maximum values of the washout parameter increased, which serves to diminish the production of asymmetry by enhancing the washout effect.

\subsection{Relation between $B-L$ asymmetry and baryon asymmetry}
Finally, the generated $Y^q_{B-L}$ can be converted to baryon asymmetry $Y^q_B$ through electroweak sphaleron processes. In the expanding universe reactions characterized by higher rates reach equilibrium more quickly. Indeed, at the electroweak phase transition, sphaleron processes and all chirality flip processes are in equilibrium.
So, upon taking into account the mentioned equilibrium conditions and the hypercharge neutrality condition according to the neutrality of the universe, baryon asymmetry near the electroweak phase transition obtained as~\cite{Chen:2007fv} 
\begin{align}
    \mu_B = \frac{28}{79} \mu_{B-L}.
    \label{eq:chemical-condition}
\end{align}
Note that chemical potentials are not modified in nonextensive Tsallis statistical mechanics, as chemical equilibrium is a classical thermodynamics concept and does not depend on the statistical mechanics framework used to describe the system.

Now, we want to determine the relation between the chemical potential and the generated asymmetry. For this purpose, we must note that in the early universe, as explained by standard statistical mechanics, $\frac{\mu}{T} \ll \frac{p}{T}$. 
Since this inequality is valid for q = 1, its expansion centered on q = 1 will certainly provide a good approximation. By considering up to the first order expansion of Eq.~(\ref{eq:dist}), for the values $|q-1|\ll1$ it becomes
\begin{align}
    f^{q} = \frac{1}{e^{\left(E-\mu\right)/T}+ \xi} + \frac{q-1}{2} \frac{\left[\left(E-\mu\right)/T\right]^2 e^{\left(E-\mu\right)/T}}{\left[e^{\left(E-\mu\right)/T}+\xi\right]^2}.
    \label{eq:dist-app}
\end{align}
Thus, the distributions of particles and antiparticles from Eq.~(\ref{eq:dist-app}) approximately equal to
\begin{align}
    f^q = A + B \mu + O(\mu^2),\\
    \overline{f}^q = A - B \mu + O(\mu^2),
\end{align}
where constants $A$ and $B$ are independent of $\mu$.
As the zero-order distributions are equal, they are not taken into account in $n^q_i - \overline{n}_i^q$, although the first orders are taken into account in $n^q_i - \overline{n}_i^q$. So, by definition a new constant, $C$, as a function of constant $B$ we can obtain
\begin{align}
    n^q_i - \overline{n}_i^q = C \mu.
\end{align} 
Then, by dividing the last equation by $s^q$, we can obtain a relation between the chemical potential and the asymmetry
\begin{align}
    Y^q = \frac{C}{s^q} \mu.
\end{align}
At last, one can multiply both side of Eq.~(\ref{eq:chemical-condition}) to $\frac{C}{s^q}$, and obtain
\begin{align}
    Y_{B}^q = \frac{28}{79} Y^q_{B-L}.
    \label{eq:relation-between-B-L-and-baryon-asymmetry}
\end{align}

\section{Numerical results}
\label{sec:results}
In this section, before we numerically solve the obtained evolution equations we would parameterize the Yukawa matrix in the Casas-Ibarra way for three RHNs~\cite{Casas:2001sr}
\begin{align}
    y=-iU\sqrt{m} R^T(\omega_1,\omega_2,\omega_3) \sqrt{M} \frac{\sqrt{2}}{v},
    \label{eq:Casas-Ibarra}
\end{align}
where $v=246\ \rm GeV$ denotes the Higgs expectation value.
$\rm m$ is the diagonal mass matrix of the light neutrino and $U$ is the unitary neutrino mixing matrix known as the PMNS matrix (Pontecorvo-Maki-Nakagawa-Sakata matrix). The PMNS parameters and mass splits of active neutrinos considering the normal hierarchy of the masses are taken from NuFIT 5.2 with Super-Kamiokande atmospheric data~\cite{Esteban:2020cvm}. Although there is a substantial error in the Dirac phase, we choose the NuFIT reported best fit, which is $232\degree$.
Note that there are also two phases of Majorana: $\alpha_{21}$ and $\alpha_{31}$, which can take values between $0$ and $4\pi$. Here, we neglect the Majorana phases as there is no experimental approach to determine these.

Moreover, $M$ is a diagonal matrix of the RHNs masses.
Finally, $R(\omega_1,\omega_2,\omega_3)$ is a generic three-dimensional orthogonal complex matrix generated via three complex angles $\omega_i \equiv x_i+iy_i$.
It is expressed in the following form like  the PMNS matrix,
\begin{align}
    R=
    \begin{pmatrix}
        1&0&0\\0&c_{\omega_1}&s_{\omega_1}\\0&-s_{\omega_1}&c_{\omega_1}
    \end{pmatrix}
    \begin{pmatrix}
	c_{\omega_2}&0&s_{\omega_2}\\0&1&0\\-s_{\omega_2}&0&c_{\omega_2}
    \end{pmatrix}
    \begin{pmatrix}
	c_{\omega_3}&s_{\omega_3}&0\\-s_{\omega_3}&c_{\omega_3}&0\\0&0&1
    \end{pmatrix},
    \label{eq:R-matrix}
\end{align}
where $c_{\omega_i} = \cos \omega_i$ and $s_{\omega_i} = \sin \omega_i$.
In summary, there are ten free parameters in the theory. They are listed in Tab. \ref{tab:parameters} along with their values considered in this work.
\begin{table}[h]
    \caption{Free parameters of the theory: $m$ is the lightest active neutrino mass, $M_i$ are RHNs masses, $x_i$ and $y_i$ are parameters of the $R$ orthogonal complex matrix \label{tab:parameters}}
    \begin{ruledtabular}
        \begin{tabular}{c c c c c c c c c c} 
            $m/{\rm GeV}$ & $M_1/{\rm GeV}$ & $M_2/{\rm GeV}$	& $M_3/{\rm GeV}$ & $x_1/\degree$ & $y_1/\degree$ & $x_2/\degree$ & $y_2/\degree$ & $x_3/\degree$ &   $y_3/\degree$ \\
            \colrule
            $10^{-11}$	& $10^{11}$ & $10^{11.6}$& $10^{12}$ & $12$ & $51.4$ & $33$ & $11.4$ & $180$ & $11$\\
        \end{tabular}
    \end{ruledtabular}
\end{table}

\begin{figure}[h]
    \includegraphics[width=140mm]{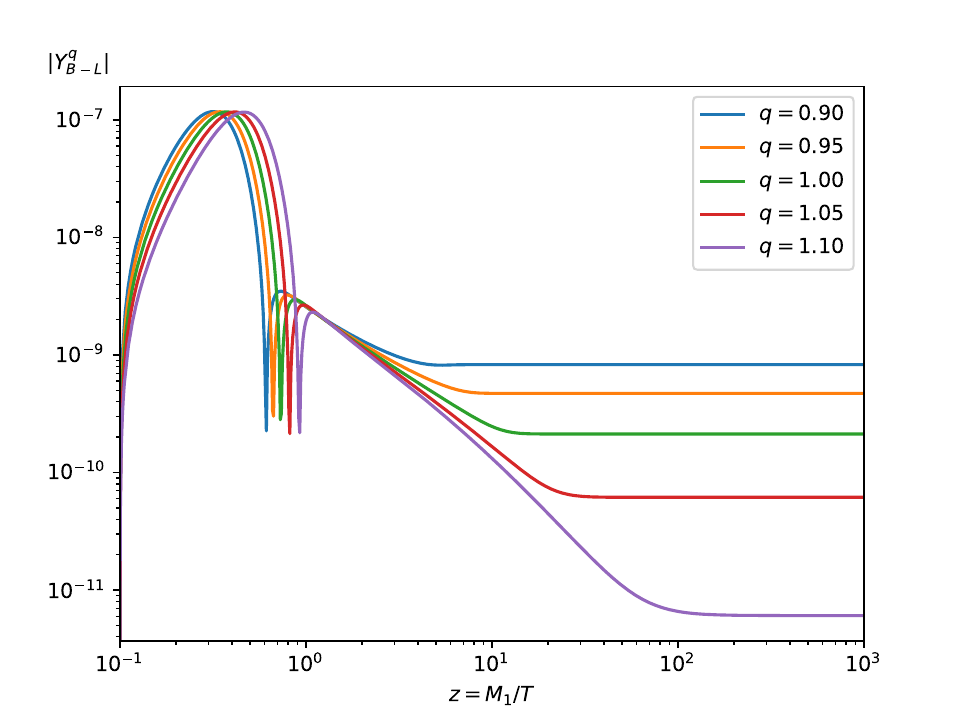}
    \caption{Evolution of $|Y^q_{B-L}|$ for some $q$ values \label{fig:YBL}}
\end{figure}
\begin{figure}[h]
    \includegraphics[width=140mm]{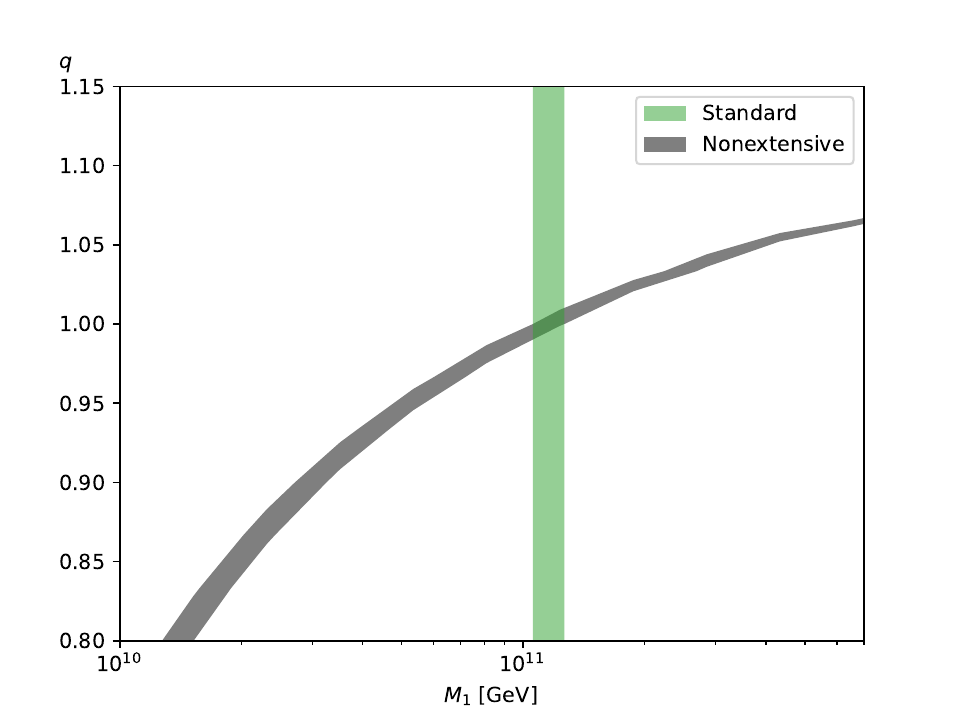}
    \caption{Valid region for $q$ and $M_1$ parameter space with $5\%$ deviation from $Y^{\rm obs}_{B}$ \label{fig:YB}}
\end{figure}

Now, we want to numerically solve the obtained evolution equations simultaneously from the starting point $z_0=10^{-1}$ to the electroweak phase transition with zero initial asymmetries. By solving the evolution equations, we present the numerical solutions of $Y_{B-L}^q$ in Fig.\ \ref{fig:YBL}. The initial conditions for the different values of $q$ are shown in the figure. 
Through the selection of this parameter space, it has been determined that for values of $q<1.2$, we are in the strong washout regime, characterized by $\Gamma_1 > H^q(T = M_1)$. In this regime, the result is independent of $Y^q_{N_1}(z_0)$~\cite{Buchmuller:2004nz}. Therefore, under the assumption of $Y_{N_1}^q(z_0)=0$, Fig.\ \ref{fig:YBL} shows that compared to the standard approach, for $q<1$, the generated $Y^q_{B-L}$ is more, while for $q>1$, the generated $Y^q_{B-L}$ is less. These results are consistent with the discussion in Sect. \ref{sec:W1}.
Indeed, for $q<1$ as opposite to $q>1$, delayed production of asymmetry arises from the delayed rising of the decay parameter which is shown in Fig.~\ref{fig:D1} and washout becoming unimportance arises from the moving of maximum point of washout parameter to high energies which is shown in Fig.~\ref{fig:W1}.
Finally, we can convert the generated $Y^q_{B-L}$ near the electroweak phase transition to baryon asymmetry using Eq.~(\ref{eq:relation-between-B-L-and-baryon-asymmetry}).
In Fig. \ref{fig:YB}, we investigate the parameter space allowed $M_1$ and $q$ considering $M_2=M_1 \times 10^{0.6}$ and $M_3=M_1 \times 10^{1}$ to generate the baryon asymmetry observed in the electroweak phase transition via leptogenesis, using both standard and nonextensive statistical mechanics.
Thus, considering the values of $q<1$, one can reduce the RHN mass required to generate the expected baryon asymmetry.

\section{Conclusion}
\label{sec:conclu}
In this study, we review the impact of nonextensive Tsallis statistical mechanics on cosmology and then examine its effect on thermal leptogenesis.
The study found that the nonextensive Tsallis statistical mechanics can affect the production of baryon asymmetry in thermal leptogenesis by modifying the equilibrium abundance of particles, decay, and washout parameters. Indeed, we indicate that the washout parameter for $q<1$ cases is feeble, so the CP violation is stronger and the generated baryon asymmetry is greater than the standard one. We also demonstrate that the numerical solution of the evolution equations agrees with this argument.
Finally, we emphasize that the RHN mass scale can be reduced with $q<1$ values. In contrast, if future works can constrain $q>1$, thermal leptogenesis requires a larger $M_1$ than the standard.

Using nonextensive Tsallis statistical mechanics in other baryogenesis scenarios, especially modern leptogenesis scenarios in future studies may offer new insights into baryon asymmetry production.

\acknowledgments
The author is very grateful to S. Safari, S. S. Gousheh, and G. Lambiase for their useful comments.

\bibliography{biblio}

\end{document}